\documentclass{aa}  
\usepackage{graphicx}
\usepackage{txfonts}
\def\beq{\begin{equation}}
\def\eeq{\end{equation}}
\def\bea{\begin{eqnarray}}
\def\eea{\end{eqnarray}} 

\begin{document}
\title{A New Approach to the Optimal Target Selection Problem}
   
\author{E. C. Elson
       \inst{1,2}
       \and
       B. A. Bassett
       \inst{1,2}
       \and
       K. van der Heyden
       \inst{1}
       \and
       Z. Z. Vilakazi
       \inst{3,4}
       }

\institute{South African Astronomical Observatory, Observatory, Cape Town, South Africa\\
          \email{ed@saao.ac.za, bruce@saao.ac.za, heyden@saao.ac.za}
          \and
          Mathematics Department, University of Cape Town, Rondebosch, South Africa\\
          \and
	  Department of Physics, University of Cape Town, Rondebosch, South Africa\\
	  \email{vilakazi@naledi.phy.uct.ac.za}
          \and
	  iThemba Labs, Faure, Cape Town, South Africa
	  }

   
\abstract
{This paper addresses a common problem in astronomy and cosmology: To optimally select a subset of targets from a larger catalog.  A specific example is the selection of targets from an imaging survey for multi-object spectrographic follow-up.} 
{We present a new heuristic optimisation algorithm, HYBRID, for this purpose and undertake detailed studies of its performance.} 
{HYBRID combines elements of the simulated annealing, MCMC and particle-swarm methods and is particularly successful in cases where the survey landscape has multiple curvature or clustering scales.}
{HYBRID consistently outperforms the other methods, especially in high-dimensionality spaces with many extrema. This means many fewer simulations must be run to reach a given performance confidence level and implies very significant advantages in solving complex or computationally expensive optimisation problems.}
{HYBRID outperforms both MCMC and SA in all cases including optimisation of high dimensional continuous surfaces indicating that HYBRID is useful far beyond the specific problem of optimal target selection.  Future work will apply HYBRID to target selection for the new 10m Southern African Large Telescope in South Africa.}


\maketitle

\section{Introduction}

In many areas of life one is faced with the problem of allocating limited resources to achieve maximal effect. In the case where this allocation takes the form of selecting a discrete subset of targets for further study we have the optimal target selection problem. As an example, the targets may be military or mining: given a large enemy fleet, which subset of targets should be attacked in order to inflict maximal damage given finite defensive capability? In the mining context, given geological and geographical information what are the optimal locations for new mine shafts/pits to be opened? Although our discussion will have an astronomy focus, the formalism we develop will be general. 

Classic examples of target selection already implemented were the spectroscopic 2df (Colless et al. 2001) and Sloan Digital Sky Survey (SDSS) galaxy surveys. In the case of the SDSS, two samples of galaxies were selected: the main sample (Strauss et al. 2002) and the luminous red galaxy (LRG) (Eisenstein et al. 2001) sample. Both were selected with an emphasis on uniformity, providing flux and volume limited samples respectively. 

Next-generation Baryon Acoustic Oscillation (BAO) surveys such as KAOS/WFMOS (Bassett et al. 2006) will take spectra for over two million galaxies at redshifts ranging from $z \sim 1$ to $z \sim 2.5-4$ over areas exceeding 500 deg$^2$. These target galaxies will have to be carefully selected from multi-colour imaging surveys and a key question will be the extent to which the survey will trade off generality for gain in addressing specific questions (e.g. the dynamics of dark energy) (Blake et al. 2006). 

Optimal target selection can have many facets. For example, cosmological errors often comprise two terms: cosmic variance and shot noise. The first pushes the survey to large areas and volumes while minimising short noise pushes the survey to smaller areas and higher densities (at constant total survey time). Area can often be obtained ``free of charge" by sparsely sampling the sky, i.e. with an inhomogeneous fibre density over the sky (Blake et al. 2006). Of course, this has to be folded into the actual realisation of the galaxy population and in general the optimal choice of targets must be done together with the optimisation of the general parameters of the survey (Bassett 2005).

In this paper we will focus on a different target selection problem in which we want to choose targets in a relatively small number of fields of view (FoV) for maximal effect, as opposed to for example selecting the optimal type of galaxy (e.g. blue or red sequence etc...). The problem we address is typical for smaller surveys where one can `cherry-pick' the best FoV for study. An example in this category is the gathering of pairs of Lyman-$\alpha$ spectra of quasars to obtain constraints on dark energy via the Alcock-Paczynski test, see e.g. McDonald, Miralda-Escude 1999. 

Section (\ref{gen}) discusses the general formalism we will be using while sections (\ref{find}) and (\ref{results}) present the HYBRID algorithm and results respectively. In this paper we extensively use the acronyms FoM and FoV standing respectively for `figure of merit' and `field(s) of view'. 

\section{General formalism}\label{gen}

As with all optimisation we need a Figure of Merit (FoM, also known as the utility, cost or objective function) which gives us an indication of the suitability/desireability of a given scenario. By maximising or minimising this FoM (as appropriate) we therefore select the best scenario for the problem at hand. To be concrete and without loss of generality, we will consider maximisation of the FoM as our goal.

Optimal target selection typically depends on various inter-related issues:
\begin{itemize}
\item The nature of the instrument that will be used to undertake the survey (e.g. the size of the telescope, the size of the field of view etc...).
\item The nature of the input target catalog (how many dimensions does it span?, how many objects does it contain? etc...)
\item The nature of the constraints (what is fixed: total time?, total cost? etc...).
\end{itemize}

A crucial, but somewhat hidden, role is played by the size and shape of the Field of View (FoV). In general it can have any shape and size but in this paper is always taken as circular of radius $R$. Two key dimensionless parameters which determine what method should be used are the ratios $L/R$ and $R_0/R$, where $L$ is the characteristic size of the input catalog (on the sky) and $R_0$ is the characteristic clustering scale of the data on the sky. 

\subsection{The discrete case}

Consider a large but discretely distributed input catalog of targets, T. In astronomical applications a classic example is a collection of galaxies. Objects in T will differ in spatial position (x, y, z) or (RA, DEC, z). In addition, targets may carry extra information, such as their colours (e.g. in the SDSS survey one has $u, r, i, g, z$ colours), discrete information regarding type (QSO, LRG, spiral, elliptical etc...) and so on.  In general elements of T will be $n$-dimensional vectors. Some of the components of these vectors can change continuously while others are discrete. 

The most basic approach to optimal target selection is to maximise the following Figure of Merit (FoM) constructed by summing over the FoM of each field of view:
\begin{equation}
\mbox{FoM} = \sum_i \parallel \mbox{FoM}_i \parallel  = \sum_{ij} w_{ij} 
\label{fom}
\end{equation}
where the sum over $i$ denotes the sum over all fields of view (FoV) in the survey, the sum over $j$ denotes objects in
the $i$th FoV. Each object is given a weighting $w_{ij}$ which determines how useful it is to the overall survey aims. Since we do not want to double-count any objects this must be taken into account in the computation of the FoM (often optimal FoV may overlap to some extent). This requirement is denoted with $\parallel \cdot \parallel$. The FoM must be optimised subject to a constraint such as:
\begin{equation}
\sum_i t_i \leq t^* \,.
\label{constraint}
\end{equation}
Here $t^*$ could represent the total time available for the survey, in which case the total number of fields to be observed is not necessarily constant (some FoV may contain brighter targets than others). The constraint may also be of a simpler form such as the total number of FoV being fixed (the case we consider in this paper). 

The choice of weightings, $w_{ij}$, will depend on the input catalog and aims of the survey. As an extreme example, galaxy surveys will give zero weight to stars in the same FoV, indeed stars may even be given negative $w_{ij}$ to discourage looking through the galactic plane. A more interesting example is that $w_{ij}$ may be chosen to implement hard or soft colour or redshift cuts or be designed to minimise selection or other biases.

In our discrete examples below, we have chosen 2-dimensional target catalogs, but the ideas scale trivially to higher-dimensional catalogs. Note that in general, T can have both a continuous and discrete dependence. 

\subsection{The continuous case}

In general, the input catalog may effectively be continuous. For example, we may have an effectively continuous map such as a CMB or X-ray map rather than a discrete set of sources. While all maps are  essentially discrete at the pixelisation level of the detector, there may be a huge number of pixels in the field of view or one may simply be dealing with a time-ordered stream of data.

In this case, it is more appropriate to define the FoM to be 
\begin{equation}
\mbox{FoM} = \sum_i \left(\int_{\Omega_i} W(\theta,\phi) d\theta d\phi\right)
\label{contfom}
\end{equation}
where $\Omega_i$ defines the interior of the $i$th FoV (and ensures the same region is not double counted) and $\theta,\phi$ parametrise position on the sky. The weight function $W$ is now a continuous function of angle on the sky. Again the FoM would be optimised subject to a constraint, typically that the total amount of survey time is fixed.

A relevant example is given by spectroscopic followup of clusters discovered using cosmic microwave background maps sensitive to the Sunyaev-Zel'dovich effect, such as will be the case for ACT (http://www.hep.upenn.edu/act/), SPT (http://spt.uchicago.edu/) and similar surveys. In this case, $W(\theta,\phi)$ could be chosen to be the temperature decrement or some other useful quantity measuring the significance of the detection or mass of the cluster. 

\subsection{Cross-correlating multiple data sets}

Optimal target selection comes into its own when it is used on multiple data sets (where human abilities start to fade). For example, given a limited spectroscopic subsample of bright galaxies, one may combine it with deep photometric images of the same part of the sky to search for bright galaxies preferentially in the middle of clusters and hence surrounded by a large number of less bright galaxies. This would provide additional targets for multi-object spectroscopy.

One may combine multiple data sets in several ways. First one could compute the cross-correlation between the various data sets in each FoV. Alternatively, if one is simply trying to maximise the effective number of targets, one can extend Eq. (\ref{fom}) by summing over the corresponding objects in each data set visible in that FoV. As an example, there may be only 3 objects visible in a given FoV in the first catalog but perhaps 10 are visible in the same FoV in the second catalog. 

Of course, the eventual aim is typically to minimise errors on an estimate of some relevant quantity or set of parameters. Hence the extra objects may not be worth while in the sense that they may be very faint and require a great deal more integration time to acquire. This can be dealt with by making the weights, $w_{ij}$ depend on the brightness or redshifts of the objects. A totally integrated approach to this problem would make use of a framework such as the Integrated Parameter Space Optimisation (IPSO) formalism (Bassett 2005) which optimises the design of a survey by using a FoM that depends directly on the size of the final error bars on quantities of interest. Here we neglect this final step for simplicity. 

\section{Finding the optimal targets}\label{find}

Having defined our FoM, we now have a function defined, in the simplest cases, over the subset of the sky defined by the target catalog. The aim is now to find the set of FoV that maximise the FoM while respecting the constraint (\ref{constraint}). This presents an unusual global optimisation problem. The aim is to find a group of directions for all the FoV which, {\em taken together}, maximise the FoM. 

In general this is a non-trivial problem since it is non-local.   One possible approach is to use grid division:  divide the target catalog on the sky into a lattice and compute the number of objects within each grid element.  The best survey is then trivial: simply strategically place a FoV in each of the N most densly populated grid elements (assuming the FoV are small enough that neighbouring FoV do not overlap).  

This method is flawed in several ways, however.  Firstly, the optimal number of grid elements to use is not obvious, it depends very much on the particular data set.  A finer grid mesh is not always better.  It might be the case that for a particular grid size, a cluster small enough to fit into one grid element is shared evenly between a few grid elements.  The over-density of this cluster will then not be as obvious as it would be if it were fortunate enough to lie only in one grid element.  Secondly, the computation time is proportional to the number of grid elements and usually proportional to the square of the number of objects and hence is unfeasible in general for very large data sets (which is the case we are mainly interested in).

When the data are sparsely distributed a much more efficient approach is to center a FoV on each point and then compute the best survey that way. This again is inefficient for large data sets and suffers from the problem that the best FoV will rarely be centered on one of the data points. 

An improvement to the lattice approach is to use an adaptive grid which iteratively refines the grid in the best areas. 
We did not persue this approach because of the added complexity. Instead we concentrated on heuristic and stochastic methods. 

\subsection{A new search algorithm - HYBRID}

To address the problem of optimisation in general, and optimal target selection in particular, we have designed a new 
heuristic algorithm we call HYBRID, which combines elements of Simulated Annealing (SA) (Kirkpatrick et al. 1983), Markov-Chain Monte Carlo (MCMC) (Metropolis et al. 1953) and Particle Swarm Optimisation (PSO) heuristics. HYBRID is a stochastic search algorithm whose basic idea is to run $m$ FoV simultaneously on the data set (where $m$ is determined by the number of FoV to be observed). At each step in the simulation, information about the performance of each FoV (encoded in its own FoM, denoted FoM$_i$) is shared among the $m$ FoV and the information is used to guide the future dynamics of each FoV. In this sense both MCMC and SA are special cases of HYBRID in which no information is shared between different FoV.  

While this key idea in HYBRID can be implemented in many ways, we have chosen the following implementation.
As with standard MCMC and SA implementations, each FoV moves randomly around the allowed region according to the law:
\begin{equation}
{\bf x}_{j+1} = {\bf x}_j + {\bf \xi}_j
\end{equation}
where ${\bf \xi}_j$ is drawn from an appropriate multivariate probability distribution. Typically this is chosen to be a multivariate Gaussian with variance ${\bf \sigma}_j$. In the case of 
standard MCMC, ${\bf \sigma}_j$ is the same at all steps while for SA the acceptance probability of a bad step decreases monotonically with step number $j$ although the variance ${\bf \sigma}_j$ is constant. Corresponding to each ${\bf x}_j$ there is a FoV with Figure of Merit (FoM) labelled FoM$_{i,j}$, i.e. 
step $j$ of the $i-th$ FoV. The proposed new position ${\bf x}_{j+1}$ is accepted with a probability governed by the Hastings-Metropolis prescription, i.e. with
probability:
\beq
P({\bf x}_{j+1}|{\bf x}_j) = \mbox{min}(e^{\alpha (FoM_{i,j+1}-FoM_{i,j})},1)
\eeq
In other words, if FoM$_{i,j+1}$ $>$ FoM$_{i,j}$ the step is always accepted (here we assume the aim is to maximise the FoM), otherwise the system accepts a worse step (FoM$_{i,j+1} \le $FoM$_{i,j}$) with a reduced probability that depends on $\alpha$ which controls how lenient the algorithm is towards accepting worse steps.

For the HYBRID algorithm we split at any step $j$, the $i$th component of the vector variance ${\bf \sigma}_j$ into the product:
\beq
\sigma^i(j) = \sigma_i^0 \times f_i(j) \times g(j)\,.
\label{sig}
\eeq
In other words the variance of the probability distribution from which the step size for a particular FoV is drawn at each step depends on two functions, $f$ and $g$ as well as an overall constant normalisation $\sigma_i^0$, that can be different in the different dimensions of the space (we typically took it to be a fixed fraction of the input catalog size in each dimension).

The function $f_i$, which we call the penalty function, shares information spatially among the various FoV. 
The basic idea is simple: if a given FoV is doing very badly relative to the other $m-1$ FoV, it is in a
bad region and statistically speaking should take bigger steps to get to a better region. If the field of view is doing very well relative to the rest of the FoV, it should take very small steps and probe its immediate neighbourhood for even better configurations. 

This deals with the annoying habit of optimal target selection that while one or more FoV may quickly find a rich cluster, they will typically wander off before the other FoV can find good regions and hence the final ``optimal" survey will typically be significantly sub-optimal. 

\begin{figure}[!ht]
\centering
\includegraphics[angle=0,width=.48\textwidth]{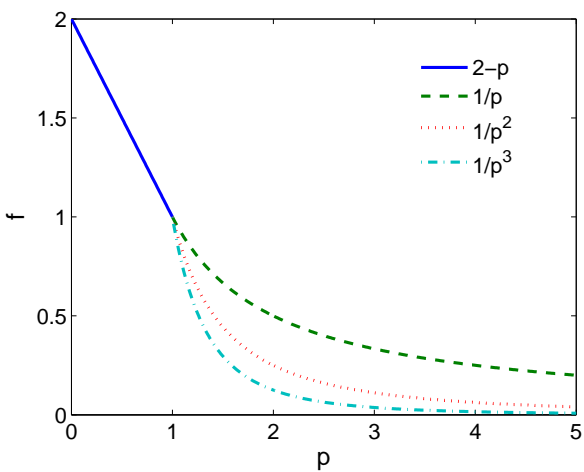}
\caption{\label{f}Various forms of the function $f$ used in our simulations as a function of $p \equiv $FoM$_i(j)/\langle $FoM$(j) \rangle$. $f$ enforces the requirement that the FoV take small steps in rich areas ($p \gg 1$) and large steps in poor areas ($p \ll 1$).}
\end{figure}

We let $f_i$ depend monotonically on the variable $p_i(j) \equiv $FoM$_i(j)/\langle $FoM$(j) \rangle$, 
the ratio of the FoM of the $i$th FoV to the average FoM (both at step $j$). In our simulations we chose $f(p_i)$ to 
be linearly decreasing from $p_i=0$ to $p_i=1$ and decreasing as $p^{-\gamma}$ for $p_i > 1$, typically with $\gamma = 2$, and the matching condition $f(p_i = 1) = 1$, as shown in Fig. (\ref{f}).  

When the FoM is continuous and $p \gg 1$ ($\ll 1$ for minimisation) it can be very profitable to 
implement a transition in $f$. When $p \gg 1$ the FoV is performing very well and hence one can be fairly confident 
of the FoV being near a local maximum and hence one can adapt $f$ to the natural scale of that maximum, namely one can 
make a rough estimate of the curvature of the extremum by histograming the recent chain values in each dimension. 
One can then choose $f$ to be the standard deviation of this histogram in each dimension. Hence the characteristic step taken will be adapted to the local curvature of the extremum. A similar version to this was used in our tests of Griewangk's function in section (\ref{griewangk}) and greatly improved the final approach to the minimum.  

\begin{figure}[!ht]
\centering
\includegraphics[angle=0,width=.48\textwidth]{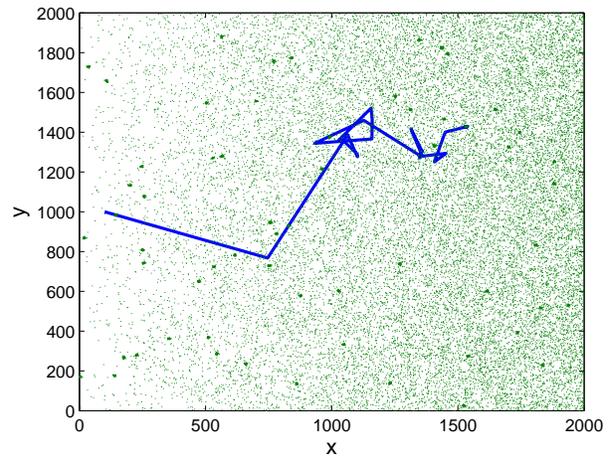}
\caption{Path of one FoV (from left to right) across the gradient+clusters data-set described in section (\ref{grad-section}). At this scale the path appears to end abruptly two-thirds the way across the dataset. Fig. (\ref{pzoom}) shows how the FoV converges rapidly onto a rich cluster.}
\label{path}
\end{figure}

\begin{figure}[!ht]
\centering
\includegraphics[angle=0,width=.48\textwidth]{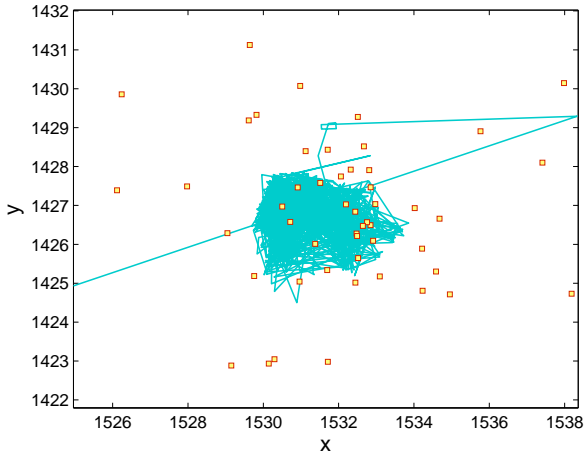}
\caption{Zoom of Fig. (\ref{path}) showing the final part of the trajectory of the FoV centre after locking onto a cluster. On finding the cluster the average step size plummets by a factor $\sim 10^{-3}$ allowing the FoV to fully explore the cluster. With standard MCMC or SA methods the FoV would have evolved away from the cluster before exploring it.  Cluster points are denoted by squares. On average, away from clusters, a region this size would contain less than 
one point. The radius of the FoV for this data set is 5 showing how the optimal FoV captures most of the points in the cluster.}
\label{pzoom}
\end{figure}

\begin{figure}[!ht]
\centering
\includegraphics[angle=0,width=.48\textwidth]{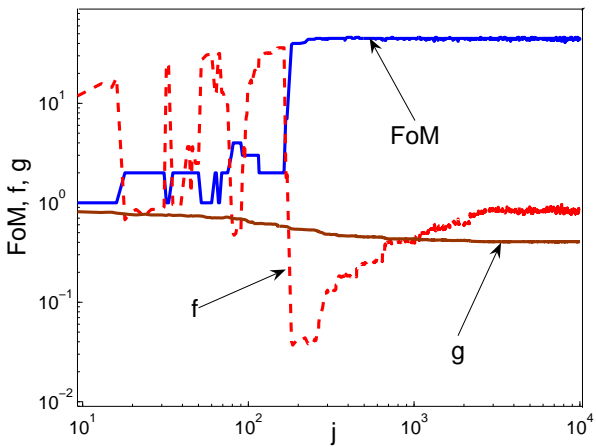}
\caption{Behaviour of the FoM for the single FoV shown in Figs (\ref{path}) and (\ref{pzoom}) as well as the corresponding $f$ and $g$ as a function of step number $j$.  Note how $f$ varies rapidly, anti-correlated with the FoM while $g$ slowly decreases so as to relax the full system of FoV towards the global optimum. At $j \sim 150$ the FoV comes across the cluster and explores it to $j > 10^4$ illustrating how effective HYBRID is. The increase of $f$ for $j > 200$ shows the improved performance of the remaining FoV in the survey which asymptotically match the performance of 
this FoV (since $f \rightarrow 1$).}
\label{one-FoV}
\end{figure}

A different approach in this situation is to switch to a purely deterministic method. When $p$ is sufficiently 
large one can make a deterministic estimate of the gradient and use any of the standard gradient-based methods to converge to the extremum analytically, basically by approximating the FoM as quadratic near the minimum. In summary, the function $f$ implements the spatial and parallel sharing of information as occurs in particle swarm 
optimisation and ensures that good performers are rewarded while poor performers are aided. 

On the other hand, the function $g$ is chosen to help the whole set of FoV settle into their local optima as time goes by, thereby helping to yield an optimal total survey. The function $g$ represents a performance-dependent cooling of the average step size. 

In SA, the probability with which bad steps are accepted is reduced gradually and monotonically with time according to some cooling schedule, 
e.g. $\alpha \propto (\log(1+j))^{-1}$ (Kirkpatrick et al. 1983).  With a sufficiently slow cooling schedule one is assured of 
reaching the global optimum. However, this is slow. A faster cooling schedule may deliver better results 
in some cases but may also be trapped in glassy, local extrema - there is no assurance that the global optimum
will be found.  

Instead of letting $g$ depend directly on $j$, we again share information between the $m$ FoV. We let $g$ depend 
on the variable $q \equiv \langle $FoM$(j) \rangle/\langle $FoM$(1) \rangle$, i.e. the ratio of the average FoM at step 
$j$ to that at step $1$. Hence, the evolution of the $g(q)$ may be non-monotonic and there is no need for reannealing (increasing $g$ by hand at some point) - if the survey is performing badly the step size will automatically increase. In our simulations we chose:
\beq
g(q) = q^{-\beta}
\eeq
with $\beta \in [0.1,1]$. For large $q$ the survey is performing well and $g$ is small. 

Figures ($\ref{path}$, $\ref{pzoom}$, $\ref{one-FoV}$) show the evolution of a single FoV through the gradient dataset described in section (\ref{grad-section}) which ended-up locking on to a cluster. The discovery of a rich cluster causes $f$ to drop by a factor $10^3$ allowing the FoV to stick to the cluster and explore it thoroughly.  Figure ($\ref{one-FoV}$) shows that after a few hundred steps the FoM begins to level off.  Notice, however, that thereafter, even though the FoM remains more or less constant, the magnitude of $f$ increases by about a factor of 20 so that after 10000 steps it's back up to 1.  The reason for this is that while the FoV is exploring the cluster, the other $m-1$ FoV are themselves moving to progressively better regions.  This has the effect of diminishing the performance of the FoV relative to the others and so $f\rightarrow$ 1.  The FoV should still scout the cluster even though its performance relative to the other FoV is deteriorating.  In order to ensure this, the cooling schedule, controlled by $g$, should decrease the step sizes of all FoV.  Thus, even though the FoV, after several hundred steps, is on a par with the other $m-1$ FoV, all the FoV are taking smaller steps thanks to $g$.  By handing over control of the FoV to $g$ at later times, each FoV can still effectively explore optimal regions by being forced to take smaller step sizes.

The third component to HYBRID is the acceptance ratio, $\alpha$. In the standard Hastings-Metropolis method for MCMC, 
a jump to a worse position is accepted with probability $\exp(\alpha\Delta)$ where $\alpha = 1/2$ and $\Delta = \mbox{FoM}_{j+1} - \mbox{FoM}_j$ is the difference in FoM between the proposed new step and the current step. Clearly $\alpha$ plays a key role in determining how likely the system is to accept a bad step (which is important in escaping from local minima). 

In HYBRID we can assign a different value of $\alpha$ to each FoV and to depend on step $j$ as well as allowing the resulting 
$\alpha_i$ to depend on the information gathered by the system, as with the variance $\sigma_j$. The basic philosophy for this is as 
follows: if a FoV quickly finds a good region of targets we do not want it wandering off while the rest of the FoV 
search for greener pastures. One could ensure this by fixing $\alpha$ to be large, but then FoV would get stuck 
even in relatively poor regions and not be able to look for anything better. Conversely, if $\alpha$ is small, the FoV 
will leave excellent regions before the rest of the FoV find good regions.

Therefore allowing $\alpha$, like $f$ and $g$ to depend on the ratio $p$ provides a dynamic way of helping FoV to ``stick" to good clusters. Here ``good'' is defined relative to the average FoM of all the various FoV, i.e. by $p$. A simple choice therefore for the functional form of $\alpha_i(p)$ is to make it roughly go as $f^{-1}_i(p)$.
However, in our simulations below we choose $\alpha$ constant for simplicity (except in the case of simulated annealing).

The above choices for $f, g$ and $\alpha$ are motivated by the required asymptotics but are otherwise 
fairly arbitrary. It would be ideal to have a method of finding the converging automatically to appropriate functions for each data set, a problem left for future work. 
{\subsection{Simulated Annealing}

Simulated annealing corresponds to the specific case $f=g=1$.  True SA corresponds to the choice $\alpha \propto (\ln(1+j))^{-1}$, i.e. a logarithmic cooling schedule, where $\alpha$ is interpreted as the temperature of the system. This cooling ensures finding the global minimum in the infinite time limit. However, we also consider inverse power-law dependence on $j$ which typically gives better results after a relatively small number of steps. 

The problem with SA is that the acceptance probability of bad steps can become extremely small early on in the simulation even when the system is trapped in a poor position. This typically means the system must be reannealed by resetting the cooling schedule $T$ to 
some previous value. With HYBRID this is not necessary since the cooling schedule depends on step number $j$ as well as the controlling function $g$ and so reannealing occurs automatically and only when necessary.

\subsection{Step cooling}

An alternative to SA is to keep $\alpha$ constant but to slowly cool the average size of the steps, i.e. to have $g = g(j)$. We call this {\em step cooling} (SC) and have explicitly run simulations with a logarithmic step cooling schedule, i.e. $g \propto (\ln(1+j))^{-1}$. Performance on the test cases presented here was comparable to, but not quite as good, as standard SA although we do not include it in the figures for clarity. In both SA and step cooling the system ends up effectively trapped - in the former case because a smaller and smaller fraction of attempted jumps are successful and in the latter because the step size approaches zero. In all cases we tested HYBRID significantly outperforms both of these methods.

\subsection{Particle Swarm Optimisation}

Particle Swarm Optimisation (PSO) is a method modelled on natural swarms
and herds found in nature that use a large number of agents to efficiently search a
volume for an optimal point. The dynamics of each individual in the swarm depends on the
dynamics and success or failure of the other agents (see e.g. (Eberhart et al. 1995), (Kennedy et al. 2001)).  For an application to astrophysics see (Skokos et al. 2005). In the context of HYBRID PSO can naturally be implemented by setting $g = 1$ and
allowing spatial ``communication" between the FoV via $f_i(j)$.

There is an important difference between PSO as implemented in standard
optimisation algorithms and the above implementation in HYBRID. One of the
main aims of HYBRID is the study of the optimal target selection problem
where the aim is for each agent in the swarm to find the best possible
position for itself that does not overlap with other agents and which,
taken together, give the best possible combined FoM. Hence, HYBRID does
not use the positions of the FoV, only their individual performances. Of
course, this can easily be generalised in the case of standard
optimisation where the desire is to find a single optimum.}

\subsection{MCMC}

Standard Markov-Chain Monte Carlo (MCMC) is governed by a further specialisation, namely  $f=1$, $g = 1$. In this case the variance of the jump distribution is constant in time for all fields of view, as is the acceptance ratio $\alpha$. Although MCMC is not typically used as an optimisation algorithm it is assured to find the global optimum in the long-time limit and hence provides a very simple, robust optimisation method. MCMC is the blind and deaf version of HYBRID and provides a baseline which allows us to understand how useful the proposed new functions $f,g$ are.

\section{Results}\label{results}

In this section we discuss the results of the various methods (HYBRID, MCMC and SA) against various 
simulated data-sets and functions. We will show how HYBRID wins over MCMC and SA in three ways: (1) The maximum average FoM achieved is always higher. (2) The convergence to the average maximum is faster and (3) the HYBRID method is much more consistent in delivering good results so the spread around the maximum is much smaller. Combined this yields significant improvements in performance over SA and MCMC. 

In all our discrete data simulations we take $w_{ij} = 1$ and we therefore simply attempt to maximise the number of points lying within the 20 FoV which we use to define the compete survey. 

\subsection{Clusters on a uniform background}

Our first test data set consists of uniformly distributed points with multiple clusters superimposed on the 
distribution. As such it is a crude mimic of typical data sets where optimal target selection is useful, i.e. 
finding very good regions embedded in random low-density areas. While we tried many different configurations the figures
show results for 15000 points, two-thirds of which are uniformly distributed and one-third are contained in $100$ clusters of characteristic radius of $0.2\%$ the total survey size each containing 50 points. \footnote{For comparison we note that a typical galaxy cluster may subtend a few arcminutes on the sky at a redshift $z < 1$. Hence our clusters are about the appropriate size for simulating a cluster embedded in a survey covering a $50^{\circ} \times 50^{\circ}$ region on the sky.} The radius of the (circular) FoV is taken to be $0.1\%$ the size of the survey (half the characteristic radius of the clusters). 

\begin{figure}[!ht]
\centering
\includegraphics[angle=0,width=.48\textwidth]{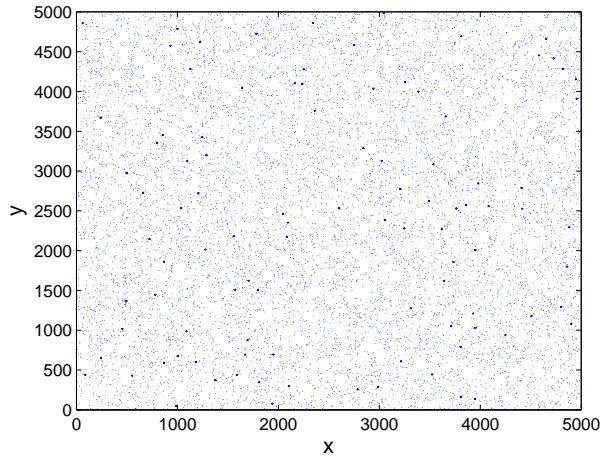}
\caption{Test dataset 1: A uniform distribution of $10^4$ points with 100 uniformly distributed dense clusters superimposed each containing $50$ points.}
\label{dataset-uniform}
\end{figure} 

\begin{figure}[!ht]
\centering
\includegraphics[angle=0,width=.43\textwidth]{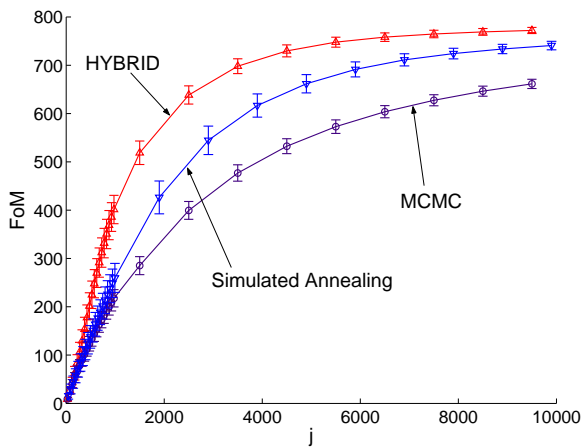}
\caption{Performance on the dataset shown in Fig. (\ref{dataset-uniform}) of the various methods as a function of step $j$ and averaged over all 5000 runs.  The errorbars show  1$\sigma$ variation in the FoM over the 5000 runs.}
\label{FoM-uniform}
\end{figure} 

The performance of the various methods is shown in Fig. (\ref{FoM-uniform}) with HYBRID outperforming 
MCMC and SA both in terms of maximum FoM and acceleration of the FoM with step number. The figure shows the 
FoM averaged over thousands of runs for each method. In addition we show the $1\sigma$ error bars on the FoM at each step. Note how for $j > 4000$ the HYBRID method has significantly smaller error bars showing that there is increased consistency as well as better performance on average. 

\subsection{Clusters on a gradient}\label{grad-section}

\begin{figure}[!ht]
\centering
\includegraphics[angle=0,width=.48\textwidth]{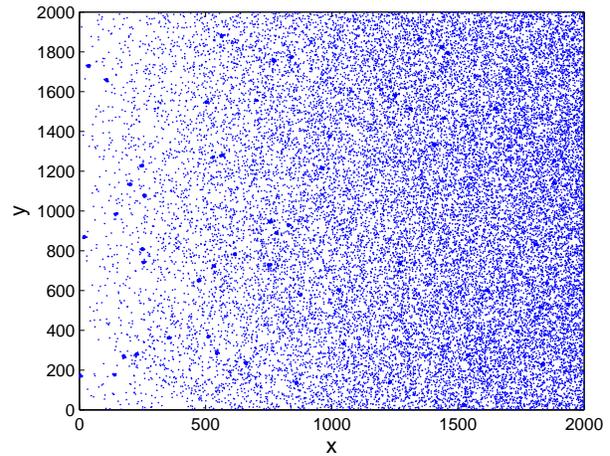}
\caption{Test dataset 2: Gradient background with 25000 points with 60 superimposed dense clusters each containing 50 points.}
\label{dataset-gradient}
\end{figure}

An extension of the first data set is to superimpose the clusters on a large gradient, shown in 
Fig. (\ref{dataset-gradient}). This dataset tests the algorithms ability to find large regions of 
high intensity as opposed to small clusters and serves as a further examination of the speed of convergence to optimal regions.

The data-set used for our simulations contained 28000 points in total with 60 embedded clusters containing 50 points each and with characteristic radius $0.5\%$ the total survey size. The radius of the FoV is taken to be $0.25\%$ the size of the survey, again half the characteristic radius of the clusters.

\begin{figure}[!ht]
\centering
\includegraphics[angle=0,width=.43\textwidth]{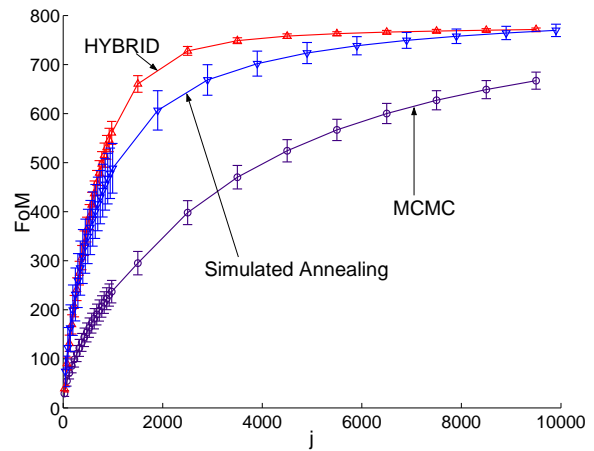}
\caption{Performance, on the gradient dataset in Fig (\ref{dataset-gradient}), of the various methods as a function of step $j$ and averaged over all runs. The 1$\sigma$ errorbars shown computed from 5000 runs.}
\label{FoM-gradient}
\end{figure} 

\begin{figure}[!ht]
\centering
\includegraphics[angle=0,width=.48\textwidth]{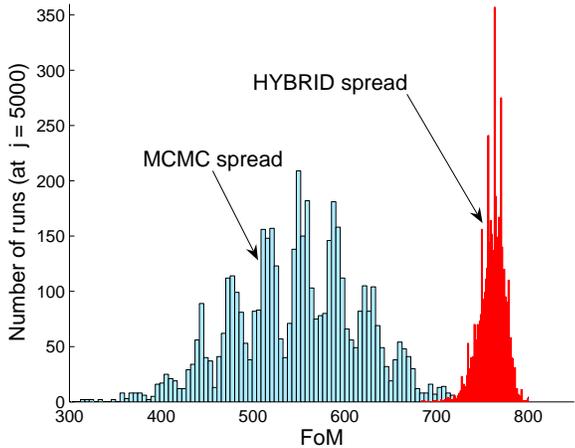}
\caption{Histogram at step $j=$5000 for the FoM for HYBRID and MCMC (SA is similar to MCMC and is not shown for clarity) on the gradient dataset of Fig (\ref{dataset-gradient}).  Notice how the HYBRID runs are significantly more clustered than MCMC runs implying much enhanced consistency: HYBRID relies much less on luck than SA or MCMC do.}
\label{histogram-gradient-5000}
\end{figure} 

The HYBRID algorithm outperforms both SA and MCMC in the same way as before, leading to higher average FoM 
and to significantly smaller error bars. This illustrates how successful the HYBRID algorithm is at sensing the 
overall gradient in the data set and responds accordingly: FoV in sparse areas are forced by the function $f$ 
to take large steps which favour moving across the gradient. 

This extra consistency is highlighted in Fig. (\ref{histogram-gradient-5000}) which shows histograms of the FoM values at 
$j=5000$ for both the HYBRID and MCMC (SA is similar to MCMC). It is clear that the HYBRID runs are significantly more
clustered together than the MCMC runs are. In practise this means that fewer HYBRID simulations need to be run to get 
good results. 

\subsection{Simulated galaxy data}

\begin{figure}[!ht]
\centering
\includegraphics[angle=0,width=.48\textwidth]{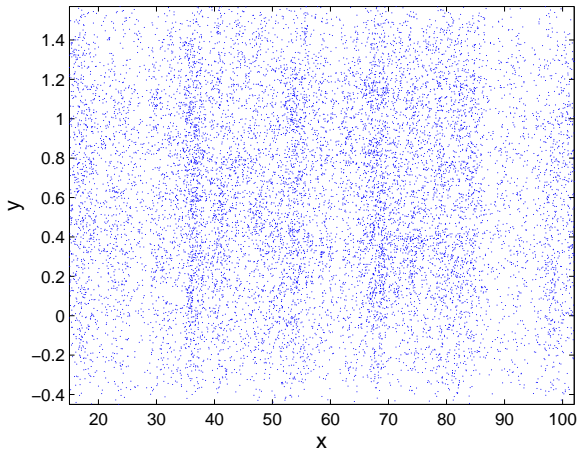}
\caption{Dataset 3: Simulated galaxy catalog with 13500 points and an inhomogeneous density along the x-axis simulating a partially completed survey.}
\label{stripy}
\end{figure}

Our final discrete data set is a simulated galaxy clustering set allowing
for an inhomogeneous input catalog. This is common in astronomy when the sky has been incompletely sampled 
leaving regions with little or no exposure and others with deep exposures and large numbers of targets. In this 
case we have mimicked a scan strategy that fills in strips at constant depth leaving the stripy pattern seen in 
Fig. (\ref{stripy}). The dataset contains 13500 points and is purposely of very different scale in the $x$ and $y$ 
directions. The radius of the FoV we use is 8 arcmin (assuming the $x$ and $y$ dimensions are RA and DEC) appropriate
to the Southern African Large Telescope. 

\begin{figure}[!ht]
\centering
\includegraphics[angle=0,width=.43\textwidth]{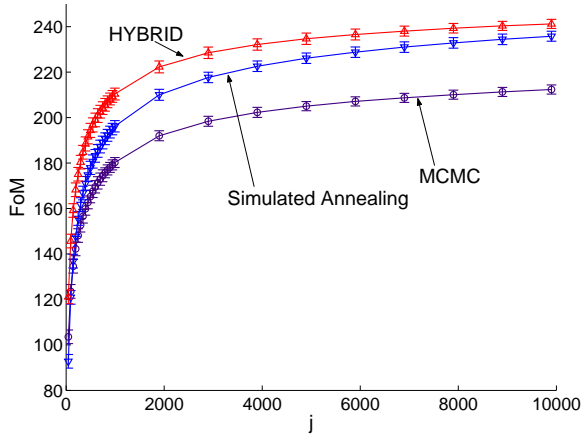}
\caption{Performance for the dataset of Fig. (\ref{stripy}) of the various methods as a function of step $j$ and averaged over all runs. The 1$\sigma$ errorbars shown were computed from 5000 runs.}
\label{2s-performance}
\end{figure} 

In this case we imposed periodic boundary conditions and choose different $\sigma^0_i$ for $i=x,y$ to allow for the elongated shape of the input space. As with the uniform and gradient datasets described above HYBRID outperforms both SA and MCMC. Although the improvement is not exponential the gains are significant and in the region of $\sim 20\%$. We now consider cases in which the improvement provided by HYBRID is much more dramatic. 

\subsection{50-dimensional hyperboloid}

To demonstrate the power of HYBRID it is useful to consider optimisation of high-dimensionality continuous functions. 
In this case the aim is simply to find a path which converges as close as possible to the global extremum of the function. Our first continuous function is a standard optimisation test-function: the $n$-dimensional hyperboloid described by (we take $n=50$):
\beq
C_1(x_i) = \Sigma_{i=1}^{50} x_i^2
\label{hyp}
\eeq
In this case the aim is to minimise the function $C_1$ which clearly occurs at $x_i = 0$ where $C_1(0) = 0$. In this case we quote the best performing path as the FoM rather than sum over all FoV \footnote{In the case of optimising continuous functions the notion of fields of view has little meaning. We are now simply computing the function at points on the hypersurface of interest.}.

\begin{figure}[!ht]
\centering
\includegraphics[angle=0,width=.43\textwidth]{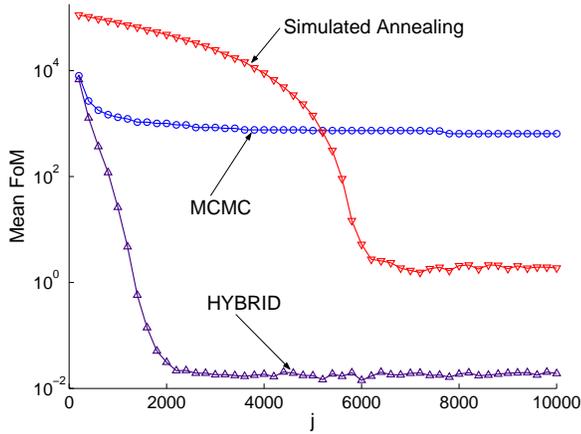}
\caption{Mean performance averaged over 5000 runs as a function of step $j$ on the continuous 50-dimensional hyperboloid given by eq. (\ref{hyp}).  HYBRID completely outperforms both SA and MCMC since the function $f_i$ allows the FoV to cascade towards the minimum by taking large steps and then to properly explore the minimum by enforcing small steps.}
\label{FoM-cont}
\end{figure} 

The performance of the various methods as a function of step $j$ and averaged over 5000 runs is shown in Fig. (\ref{FoM-cont}). HYBRID succeeds for a simple reason: FoV that have moved significantly towards the minimum will have small $f$ and hence will take small steps since they are outperforming the average FoV. Since they are taking small steps, the probability of taking a step that improves the FoM increases. Conversely FoV doing badly try to take large steps. While this is an unsuccessful strategy, on average, it works well in rare cases, leading to significant improvements. Hence the system cascades down the slope with a mixture of large, high-risk steps and small, low-risk steps. 

In contrast the MCMC method is forced either to take large steps which fail most of the time due to the high dimensionality of the system or small steps which imply a huge amount of time to reach the minimum. SA interpolates between these two by smoothly decreasing the average step size but eventually is frozen in because the step size becomes vanishingly small.

To give an idea of the performance improvements of the HYBRID algorithm, consider the best case FoM at $j=2000$ out of 5000 runs reached for each of the methods: the HYBRID, MCMC and SA algorithms 
respectively achieved $(0.03, 1006, 4.78 \times 10^4)$. Beyond $j=5500$ SA did improve significantly, surpassing the best MCMC result and reaching 
$1.874$ at $j=10,000$; an order of magnitude worse than HYBRID (1.91$\times 10^{-2}$) and taking three times longer to reach that point.  We 
do not show error bars on the points in Fig. (\ref{FoM-cont}) since now the distribution of FoM is very non-Gaussian and
cannot easily be represented on the plot since the minima in the HYBRID case can be below $10^{-10}$. 

In summary HYBRID significantly outperforms both MCMC and SA in all areas: best performance, consistency and speed to 
reach a given threshold. 

\subsection{Griewangk's test-function}\label{griewangk}

\begin{figure}[!ht]
\centering
\includegraphics[angle=0,width=.48\textwidth]{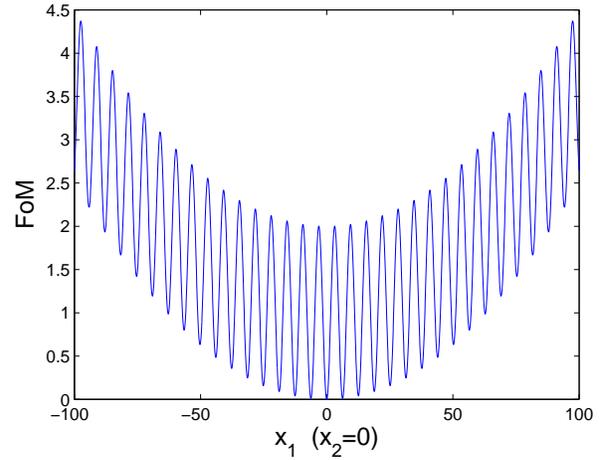}
\caption{Slice through Griewangk's function at $x_2=0$.  Note the two length scales of variability governed by the large-scale quadratic term and the small-scale oscillatory term.}
\label{grie-slice}
\end{figure}

A second continuous example is provided by another standard test-bed for optimisation algorithms: Griewangk's function. We consider only the two dimensional case which nevertheless processes a very large number of local minima as can be seen in the slice through the plane $x_2=0$ shown in Fig. (\ref{grie-slice}). The function possesses a single global minimum at $x_1^*=x_2^*=0$ where $C_2(x_1^*,x_2^*)=0$. Griewangk's function  is defined as:
\beq
C_2(x_j) = \frac{1}{4000} \Sigma_{j=1}^2 x_j^2 - \Pi_{j=1}^2 \cos\left(\frac{x_j}{\sqrt{j}}\right) + 1
\label{grie}
\eeq
where the $x_j$ can range over the interval $[-600,600]$. We choose the starting values to be $x_1=x_2 = 500$ in 
all cases. 

\begin{figure}[!ht]
\centering
\includegraphics[angle=0,width=.43\textwidth]{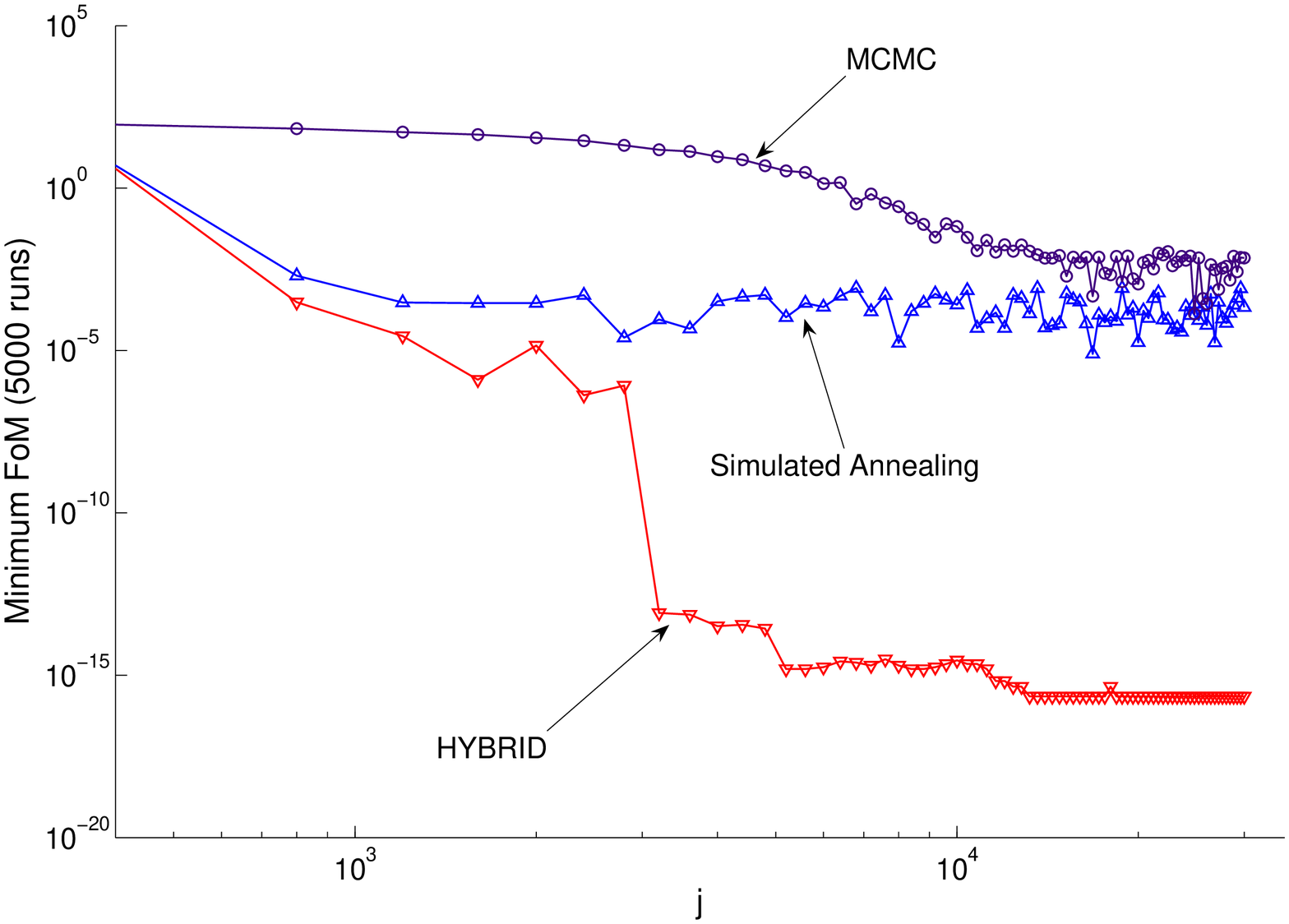}
\caption{Best performance over 5000 runs as a function of step $j$ on Griewangk's function. HYBRID significantly outperforms the other methods, reaching values of order $10^{10}$ smaller than either simulated annealing or MCMC. Each data point is the minimum FoM achieved by any FoV out of all the runs at step $j$.  For very small FoM, $f$ is made to depend only on the FoM allowing rapid convergence to very small values of the function. The best values achieved by SA or MCMC after $10^4$ steps were achieved by HYBRID after only $\sim 500$ steps.}
\label{FoM-mingrie}
\end{figure} 

Figure (\ref{FoM-mingrie}) shows the performance of the HYBRID, SA and MCMC algorithms versus step number $j$, again showing the minimum FoM of the best Markov chain (rather than the average FoM). It is clear that HYBRID completely 
outperforms both SA and MCMC. Although initially all methods lead to an exponential decrease in $C_2$, HYBRID 
has the largest exponent. Further, both MCMC and SA stagnate rapidly. Instead HYBRID continues to 
decrease exponentially, albeit at a somewhat slower rate, showing the power of the new algorithm. 

The power of HYBRID can be judged by the minimum values of the function achieved. For very small FoM, one can 
take advantage of the flexibility in $f$ to make it independent of the other FoV. For FoM $<$ $10^{-7}$ we made $f$ only 
a function of the FoM. With this modification of $f$ we were able to consistently get below $10^{-14}$ within $30000$ 
steps compared to the performance of the MCMC and SA cases which were very poor in comparison. 

HYBRID outperforms SA and MCMC despite the function only being two-dimensional for a simple reason: there are two 
curvature scales in the Griewangk function: the function varies strongly as $x$ varies by $\sim 10$ as one passes through
the many local minima. On the other hand there is an overall average parabolic shape visible as $x$ varies 
over $\sim 100$. HYBRID has the flexibility to deal with multiple curvature scales while standard methods can 
adapt to only one of the scales leading to poor performance overall. 

\section{Conclusions}

The problem of optimal target selection will be key in maximally extracting value from the large data sets 
that will characterise many sciences in the coming decades. From astronomy to genetics there is a need for a 
formalism that can optimally select targets for a specific experiment using a specific instrument. In extragalatic astronomy and cosmology there will be significant pressure to select optimal subsets of imaging data for spectroscopic followup driving the development of efficient algorithms for this purpose. 

In this paper we present a new optimisation algorithm, HYBRID, and compare its performance on simulated data using 
two standard stochastic algorithms: simulated annealing (SA) and Markov-Chain 
Monte Carlo (MCMC). The key advance in HYBRID is the idea to parallelise the search, sharing information between fields of view (`agents') at each step in the search so that each field of view has knowledge of how well it is performing relative to the ensemble of fields of view. As a result the average step size taken by a field of view can be adapted to its own performance. In this sense HYBRID is a combination of SA, MCMC and Particle Swarm Optimisation. We show how HYBRID outperforms both SA and MCMC in all cases and is particularly efficient in the cases where there are multiple clustering scales in the data and/or the target space has high dimensionality. In the case of the minimisation of known test functions, HYBRID significantly outperforms all other methods by up to ten orders of magnitude, 
indicating that HYBRID will be useful far beyond the specific problem of optimal target selection. 

Future work will apply HYBRID to target selection for the new 10m SALT telescope in South Africa.

We acknowledge use of the UCT Physics cluster, Carmen, for some of the simulations in this paper.


\begin{thebibliography}{9999}
\bibitem{2df} M. Colless {\em et al.}, MNRAS, {\bf 328}, 1039 (2001)
\bibitem{sdssmain} M. Strauss {\em et al.}, Astron.J. {\bf 124}, 1810 (2002)
\bibitem{sdsslrg} D.J. Eisenstein {\em et al.}, Astron.J. {\bf 122}, 2267 (2001)
\bibitem{wfmos}  B. A. Bassett, R. C. Nichol, D. J. Eisenstein and the WFMOS Feasibility Study Dark Energy Team, A$\&$G, October 2006, astro-ph/0510272;  K. Glazebrook and the WFMOS Feasibility Study Dark Energy Team; White paper submitted to the Dark Energy Task Force, astro-ph/0507457; http://www.noao.edu/kaos/; http://www.dsg.port.ac.uk/$\sim$bruce/kaos/.
\bibitem{chris} C. Blake {\em et al.}, Mon.Not.Roy.Astron.Soc. {\bf 365}, 255-264 (2006)
\bibitem{opti}  B. A. Bassett, Phys.Rev. D{\bf 71}, 083517 (2005)
\bibitem{ap} P. McDonald and J. Miralda-Escude, Ap. J, {\bf 518}, 24 (1999)
\bibitem{sa} S. Kirkpatrick, C.D. Gelatt, Jr., and M.P. Vecchi, Science {\bf 220}, 671 (1983).
\bibitem{mcmc} N. Metropolis, A. W. Rosenbluth, M.N. Rosenbluth, A. H. Teller and E. Teller, J. Chem. Phys.
{\bf 21}, 1087 (1953); W. K. Hastings,  Biometrika {\bf 57}, 97 (1970)
\bibitem{pso} See e.g. A. Engelbrecht, Fundamentals of Computational Swarm Intelligence, Wiley \& Sons.
\bibitem{PSO1} J. Kennedy, R. C. Eberhart, Morgan Kaufmann Publishers Inc. San Francisco, CA, USA, (2001)
\bibitem{PSO2} C. Skokos, K.E. Parsopoulos, P.A. Patsis, M.N. Vrahatis,  Mon.Not.Roy.Astron.Soc., {\bf 359}, 251-26, (2005)
\bibitem{PSO4}  R. Eberhart; J. Kennedy, Proc. IEEE Int. Conf. Neural Networks., {\bf 4}, 1942-1948, (1995)

\end{thebibliography}
\end{document}